\documentclass{elsart}
\usepackage{natbib}
\usepackage{graphicx}

\begin{document}

\begin{frontmatter}



\title{Reconnection at the Heliopause}


\author[Utrecht,Bonn]{D.H. Nickeler\corauthref{cor}},
\corauth[cor]{Corresponding author.}
\ead{D.H.Nickeler@phys.uu.nl}
\author[Bonn]{H.-J. Fahr}
\ead{hfahr@astro.uni-bonn.de}

\address[Utrecht]{Astronomical Institute, Utrecht University, Princetonplein 5, 3584 CC Utrecht, 
the Netherlands}
\address[Bonn]{Institut f\"{u}r Astrophysik und Extraterrestrische Forschung, Auf dem H\"{u}gel 71, 
53121 Bonn, Germany}

%
%
%
\begin{abstract}

 In this MHD-model of the heliosphere, we assume a Parker--type
 flow, and a Parker--type spiral magnetic field, which is
 extrapolated further downstream from the termination shock to the heliopause.  
 We raise the question whether the heliopause nose region may be leaky with respect to fields and plasmas
 due to nonideal plasma dynamics, implying a  breakdown of the magnetic barrier. We analyse some simple
 scenarios to find reconnection rates and circumstances, under which
 the heliosphere can be an "open" or a "closed" magnetosphere. 
We do not pretend to offer a complete solution for the heliosphere, on the
basis of nonideal MHD theory, but present a prescription to find such a
solution on the basis of potential fields including the knowledge of
neutral points. As an example we imitate the Parker spiral as a monopole
with a superposition of homogeneous asymptotical boundary conditions.
We use this toy model for $x<-R$ where $R=100$\,AU is the distance of the
termination shock to describe the situation in the nose region of the 
heliopause. 

\end{abstract}


\end{frontmatter}


\section{Introduction}

\begin{figure}[tbp]
\begin{center}
\includegraphics[width=6cm]{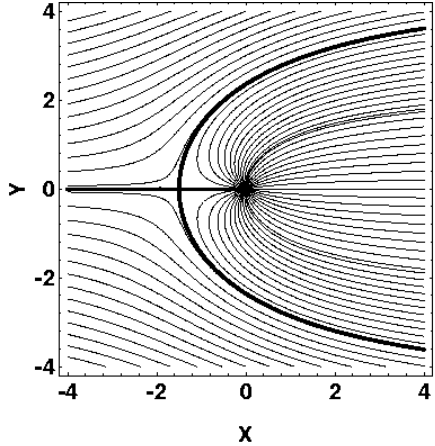}
\end{center}
\caption{The heliospheric outflow in the vicinity of the stagnation point,
similar to the classical Parker outflow. The scale is given in units of 100\,AU}
\label{parkerflow}
\end{figure}

In the past several calculations concerning the role of magnetic
reconnection in the vicinity of the heliopause have been performed. \cite
{Fahr}(and references therein) calculated reconnection probabilities and
gave estimations for reconnection rates, but without complete reconnection
solutions of the nonideal MHD equations.  Here, we present a plane model of
the heliosphere in the framework of stationary nonideal MHD. We assume that
the tail direction is the direction of the $x$--axis (see Figure \ref
{parkerflow}), which is also the direction of the asymptotical flow around
the heliosphere. The direction of invariance is the $z$--direction. Results
depend mainly on the tilt angle between the asymptotical magnetic field to the
asymptotical plasma velocity.  We concentrate on investigating the region
where the interstellar plasma encounters the solar wind plasma. 

\section{Stationary nonideal MHD--flows in 2D}

\subsection{Basic equations}

In our plane model of the heliosphere we make the following assumptions:
\begin{itemize}
\item  we take only stationary fields, i.e. $\partial /\partial t=0$, and
take: $\partial /\partial z=0$
\item  incompressibility as a substitute for the energy equation is assumed.
Incompressibility here means that the density of a fluid element moving
with the flow does not change in time meaning that the convective derivative 
$\frac{d\rho}{dt}:=\frac{\partial\rho}{\partial t} + \vec v \cdot \vec\nabla
\rho$ vanishes. Using the mass continuity $\frac{\partial\rho}{\partial t} + 
\vec\nabla\cdot (\rho\vec v) = 0$ we thus can derive with $\frac{d\rho}{dt} = 
-\rho\vec\nabla\cdot\vec v = 0$ which implies $\vec\nabla\cdot\vec v = 0$
and $\vec v \cdot \nabla\rho = 0$.
Then we introduce the auxiliary flow field or streaming vector $\vec{w}$ with 
$\vec{w}=\sqrt{\rho }\,\vec{v}\quad $leading to $\nabla \cdot \vec{w}=0$.
\end{itemize} 

Therefore, the basic MHD equations are given by the following set
\begin{eqnarray}
\vec\nabla\cdot\vec B &=& 0\\
    \vec w: &=& \sqrt{\rho}\,\vec v\\
  \vec\nabla\cdot\vec w &=& 0\\
   \vec\nabla\times\vec B &=& \mu_{0}\,\vec j\\
    -\vec\nabla\phi_{e}
    + E_{0}\vec e_{z} + \frac{1}{\sqrt{\rho}} \vec w\times\vec B &=& \vec R\label{ohmresimhd1}\\
           \vec\nabla\left( \frac{1}{2}\vec w^{2} + p \right) &=&
          \frac{1}{\mu_{0}}\left( \vec\nabla\times\vec B \right)\times
           \vec B - \left( \vec\nabla\times\vec w \right)\times\vec w
             \label{bisimhd2}
              \end{eqnarray}
where $p$ is the thermal pressure and $p+\frac{1}{2}\vec w^{2} = \Pi$ is the
Bernoulli pressure. $\phi_{e}(x,y)$ is the part of the electric potential whose 
gradient gives the poloidal field components (i.e. the components in $x$- and 
$y$-direction) of the electric field, so that
$\vec E=-\vec\nabla\phi_{e}(x,y)+ E_{0}\vec e_{z}$, and $E_{0}$ is a constant 
component of the electric field in the invariant $z$--direction. This is to
ensure that $\vec\nabla\times\vec E = -\frac{\partial \vec B}{\partial t} = 0$
(stationarity condition) and to fulfill $\frac{\partial\vec E}{\partial z} = 0$.
$\vec R$ is the nonidealness and the other symbols have their usual meaning.
Eq.\,(\ref{ohmresimhd1}) is the generalized Ohm's law (see \cite{vasyl} and 
\cite{hesse}).
For $\vec w$ and $\vec B$ we make the following ansatz:
\begin{eqnarray}
 \vec w &=& \vec\nabla\zeta\times\vec e_{z} + w_{z}\vec e_{z}\equiv\vec w_{p} +  w_{z}\vec e_{z} \\
   \vec B &=& \vec\nabla\alpha\times\vec e_{z} + B_{z}\vec e_{z}\equiv\vec B_{p} +  B_{z}\vec e_{z}
            \end{eqnarray}
where the index p on the right hand side of these equations stands for 
the poloidal part of the magnetic and the velocity field, and 
$\zeta$, $\alpha$, $B_{z}$, and $w_{z}$ are functions of $x$ and $y$.
With these assumptions we find for the generalized Ohm's law from 
Eq.\,(\ref{ohmresimhd1})
\begin{eqnarray}
  \vec w\times\vec B = (\vec\nabla\zeta\times\vec e_{z} + w_{z}\vec e_{z})\times(\vec\nabla\alpha\times\vec e_{z}
  +B_{z}\vec e_{z} )\label{gohmlaw1}
   \end{eqnarray}
which results in
\begin{equation}
 \!\!\!\!\!\!\!\!\!\!\!  -(\vec\nabla\zeta,\vec e_{z},\vec\nabla\alpha)\vec e_{z} -B_{z}\vec\nabla\zeta
   + w_{z}\vec\nabla\alpha
      = \sqrt{\rho}\left(\frac{\partial\phi_{e}}{\partial\alpha}\vec\nabla\alpha                                                                                
      + \frac{\partial\phi_{e}}{\partial\zeta}\vec\nabla\zeta + \vec R - E_{0}\vec e_{z}\right)\label{ohmresimhd2}
\end{equation}
The momentum balance equation Eq.\,(\ref{bisimhd2}) can be written as
\begin{eqnarray}
     \vec\nabla\Pi &=& -\frac{1}{\mu_{0}}\Delta\alpha\vec\nabla\alpha + \frac{1}{\mu_{0}}(\vec\nabla B_{z}\times\vec e_{z})
     \times(\vec\nabla\alpha\times\vec e_{z}) -\frac{1}{\mu_{0}} B_{z}\vec\nabla B_{z} \nonumber\\
       && + \Delta\zeta\vec\nabla\zeta + \left(\vec\nabla w_{z}\times\vec e_{z}\right)
          \times\left(\vec\nabla\zeta\times\vec e_{z}\right) + w_{z}\vec\nabla w_{z} \nonumber \\
          \nonumber\\
          &=&  -\frac{1}{\mu_{0}}\Delta\alpha\vec\nabla\alpha
              -\frac{1}{\mu_{0}}(\vec\nabla B_{z},\vec e_{z},\vec\nabla\alpha)\vec e_{z}
            -\frac{1}{\mu_{0}}B_{z}\vec\nabla B_{z}\nonumber \\
          && +\Delta\zeta\vec\nabla\zeta + (\vec\nabla w_{z},\vec e_{z},\vec\nabla\zeta)\vec e_{z}
            + w_{z}\vec\nabla w_{z}\label{eulerink}
  \end{eqnarray}
The problem here is to get rid of the $z$--component on the right hand side of equation (\ref{eulerink}), as
the $z$-dependence of the Bernoulli pressure $\Pi$ has to vanish if the problem should be restricted to a
two--dimensional problem.
This can be solved by setting
\begin{eqnarray}
 \vec 0\stackrel{!}{=} && (\vec\nabla w_{z},\vec e_{z},\vec\nabla\zeta)\vec e_{z}
 -\frac{1}{\mu_{0}} (\vec\nabla B_{z},\vec e_{z},\vec\nabla\alpha)\vec e_{z}\label{scherkomp2}\\
 \nonumber\\
  \Longrightarrow\quad \vec 0 = && \vec\nabla\zeta\times\vec\nabla w_{z} - \frac{1}{\mu_{0}}
    \vec\nabla\alpha\times\vec\nabla B_{z}\label{scherkomp3}\\
    = &&\left(\vec\nabla\zeta\times\vec\nabla\alpha\right)\frac{\partial w_{z}}{\partial\alpha} - \frac{1}{\mu_{0}}
    \left(\vec\nabla\alpha\times\vec\nabla\zeta\right)\frac{\partial B_{z}}{\partial\zeta}\label{scherkomp4}
  \end{eqnarray}
which in general then leads to
\begin{eqnarray}
 \mu_{0}\frac{\partial w_{z}}{\partial\alpha} + \frac{\partial B_{z}}{\partial\zeta} = 0 \label{scherkomp6}
   \end{eqnarray}
This is the condition for the vanishing of the $z$--component in the 
Euler-equation (\ref{eulerink}). This partial differential equation
(Eq.\,\ref{scherkomp6}) can be solved by any regular function
$S=S(\zeta,\alpha)$ which defines the shear components as
\begin{eqnarray}
  \frac{\partial S}{\partial\zeta} &=:& \, w_{z}\\
   \nonumber\\
    \frac{\partial S}{\partial\alpha} &=:& -\frac{1}{\mu_{0}}\, B_{z}\label{scherkomp5}
     \end{eqnarray}
as long as both we are dealing with a real two--dimensional problem, and there does 
not exist an implicit function of $\zeta$
and $\alpha$, or that $\zeta$ is an explicit function of $\alpha$, so that
we need to have
\begin{equation}
 \frac{\partial\left(\zeta,\alpha\right)}{\partial\left(x,y\right)}\neq 0
  \end{equation}
almost everywhere in the considered domain. Other cases has been discussed for 
example by \cite{tsin} or \cite{goed} without the constraint of 
incompressibility but with the constraint of a vanishing nonideal term $\vec R$.

It should be mentioned, that due to the mass continuity equation for the stationary and incompressible
flow the density $\rho$ is a function of the stream function $\zeta$, since
from the mass continuity equation $\vec v\cdot\vec\nabla\rho = 0$ it follows
\begin{equation}
\frac{1}{\sqrt{\rho}}\, \vec v\cdot\vec\nabla\rho\equiv
    \vec w\cdot\vec\nabla\rho\equiv (\vec\nabla\zeta\times\vec e_{z} + w_{z}
  \vec e_{z})\cdot\vec\nabla\rho = \frac{\partial\left(\rho,\zeta\right)}
{\partial\left(x,y\right)} = 0
\end{equation}
which implies that
\begin{equation}
\rho = \rho(\zeta)
\end{equation}
Considering the last steps we can rewrite Ohm's law (Eq.\,(\ref{ohmresimhd2})) 
in the form
\begin{eqnarray}
   \frac{\partial\left(\zeta,\alpha\right)}{\partial\left(x,y\right)} &=& \sqrt{\rho}\left(R^{z} - E_{0}\right)\nonumber\\
    -B_{z} w_{p} &=&  \sqrt{\rho}\left(R^{\zeta} + w_{p}\frac{\partial\phi_{e}}{\partial\zeta}\right)\nonumber\\
       w_{z} B_{p} &=& \sqrt{\rho}\left(R^{\alpha} + B_{p}\frac{\partial\phi_{e}}{\partial\alpha}\right)\nonumber\\
\label{ohmresimhd3}
     \end{eqnarray}
where $w_{p} = |\vec w_{p}|$, $B_{p} = |\vec B_{p}|$.
The first equation represents the $z$-component, and the other two equations
the $\zeta$- and $\alpha$-components (i.e. the poloidal components) of Ohm's
law.
The contravariant and normalized components of $\vec R$, which is represented by
\begin{equation}
\vec R = R^{\zeta}\left(\vec\nabla\zeta\right)^{0} +  R^{\alpha}\left(\vec\nabla\alpha\right)^{0}
         +  R^{z}\left(\vec\nabla z\right)^{0}\,,
 \end{equation}
are given by
\begin{eqnarray}
 R^{\alpha} &=& \frac{\vec R\cdot\left[\left(\vec\nabla z\right)^{0}\times\left(\vec\nabla\zeta\right)^{0}\right]}
                 { \vec e_{z}\cdot\left[\left(\vec\nabla\zeta\right)^{0}
                                 \times\left(\vec\nabla\alpha\right)^{0}\right] }
\\ \nonumber\\
R^{\zeta} &=& \frac{\vec R\cdot\left[\left(\vec\nabla\alpha\right)^{0}\times\left(\vec\nabla z\right)^{0}\right]}
                 { \vec e_{z}\cdot\left[\left(\vec\nabla\zeta\right)^{0}
                                 \times\left(\vec\nabla\alpha\right)^{0}\right] }
\\ \nonumber\\
R^{z} &=& \frac{\vec R\cdot\left[\left(\vec\nabla\zeta\right)^{0}\times\left(\vec\nabla\alpha\right)^{0}\right]}
                 { \vec e_{z}\cdot\left[\left(\vec\nabla\zeta\right)^{0}
                                 \times\left(\vec\nabla\alpha\right)^{0}\right] }
\end{eqnarray}
The superscript $0$ denotes the unit vectors in this direction.

Next we look at the Euler or momentum balance equation (Eq.\,(\ref{eulerink})):
\begin{eqnarray}
     \vec\nabla\Pi &=&  -\frac{1}{\mu_{0}}\Delta\alpha\vec\nabla\alpha
            -\frac{1}{\mu_{0}}B_{z}\vec\nabla B_{z} + \Delta\zeta\vec\nabla\zeta + w_{z}\vec\nabla w_{z}
             \label{eulerinkz}\\
\Rightarrow  \vec\nabla P &=& -\frac{1}{\mu_{0}}\Delta\alpha\vec\nabla\alpha + \Delta\zeta\vec\nabla\zeta
  \end{eqnarray}
with
\begin{equation}
 P\equiv\Pi + \frac{1}{2\mu_{0}}B_{z}^2 - \frac{1}{2}w_{z}^2 = p + \frac{B_{z}^{2}}{2\mu_{0}} + \frac{\rho}{2}(v_{x}^{2} + v_{y}^{2})
\end{equation}
This delivers the whole set of equations of motion to be solved
\begin{eqnarray}
\frac{\partial P}{\partial\zeta} &=& \Delta\zeta\qquad\qquad\,\,\,\frac{\partial S}{\partial\zeta}=w_{z}\\
\nonumber\\
-\frac{\partial P}{\partial\alpha} &=& \frac{1}{\mu_{0}}\Delta\alpha\qquad -\frac{\partial S}{\partial\alpha}=
\frac{1}{\mu_{0}}\, B_{z}
\end{eqnarray}
with two potentials, which are explicit functions of the stream function $\zeta$ and the (magnetic) flux function
$\alpha$. These equations have already been derived by \cite{neu-priest} for
the pure 2D case with $B_{z} = 0$ and $w_{z} = 0$.

\subsection{Conditions for field line conservation}

We know from \cite{vasyl} and \cite{hesse} that field line connection breaks 
down with respect to the
plasma velocity $\vec v_{P}\equiv\vec v$ if, in our stationary case the equation
\begin{equation}
  \vec\nabla\times\left(\vec v\times\vec B  \right)=\lambda\vec B\label{lcons}
    \end{equation}
for an unknown function $\lambda$, where $\vec v$ and $\vec B$ are given,
cannot be fulfilled. For solutions of non-ideal MHD we have to
find solutions with an appropriate function $\lambda$. To get reconnection solutions it is
necessary to violate condition (\ref{lcons}). Therefore, within the first step,
in order to fulfill Eq.\,(\ref{lcons})
we take Ohm's law (\ref{gohmlaw1}) and insert this into (\ref{lcons}).
\begin{eqnarray}
\!\!\!\!\!\!\lambda \vec B &\stackrel{!}{=}&
\vec\nabla\times\left[ \frac{1}{\sqrt{\rho}} \, \left( \vec\nabla\zeta,\vec\nabla\alpha,\vec e_{z}  \right) \vec e_{z}
   - \frac{1}{\sqrt{\rho}} B_{z}\vec\nabla\zeta + \frac{1}{\sqrt{\rho}} w_{z} \vec\nabla\alpha
  \right]\nonumber\\
\nonumber\\
  & = & \vec\nabla\left( \frac{1}{\sqrt{\rho}}\,\frac{\partial\left(\zeta,\alpha \right)}{\partial\left(x,y\right)}\right)
  \times\vec e_{z}
  -\vec\nabla\left(\frac{1}{\sqrt{\rho}}\, B_{z} \right)\times\vec\nabla\zeta
  +\vec\nabla\left( \frac{1}{\sqrt{\rho}}\, w_{z}  \right)\times\vec\nabla\alpha\nonumber \\
   \nonumber\\
    & = & \vec\nabla\left( \frac{1}{\sqrt{\rho}}\,\frac{\partial\left(\zeta,\alpha \right)}{\partial\left(x,y\right)}\right)
  \times\vec e_{z}
  -\left(\frac{1}{\sqrt{\rho}}\,\frac{ \partial B_{z}}{\partial\alpha}\right)\vec\nabla\alpha\times\vec\nabla\zeta \nonumber \\
 \nonumber \\
 & & +\left( \frac{1}{\sqrt{\rho}}\, \frac{\partial w_{z}}{\partial\zeta} -\frac{w_{z}}{2\rho^{\frac{3}{2}}}\,
  \frac{d\rho}{d\zeta}  \right)
  \vec\nabla\zeta\times\vec\nabla\alpha\nonumber
    \\ \nonumber \\
      & = & \vec\nabla\left( \frac{1}{\sqrt{\rho}}\,\frac{\partial\left(\zeta,\alpha \right)}{\partial\left(x,y\right)}
          \right)\times\vec e_{z}
  +\left(\frac{1}{\sqrt{\rho}}\,\frac{ \partial B_{z}}{\partial\alpha}
  + \frac{1}{\sqrt{\rho}}\, \frac{\partial w_{z}}{\partial\zeta} -\frac{w_{z}}{2\sqrt{\rho^{3}}}\,
  \frac{d\rho}{d\zeta}  \right)
  \vec\nabla\zeta\times\vec\nabla\alpha\nonumber \\
        \end{eqnarray}
For the poloidal part and the shear part $B_{z}$ of the magnetic field it 
follows immediately that 
\begin{eqnarray}
  \frac{1}{\sqrt{\rho}}\,\frac{\partial\left(\zeta,\alpha \right)}{\partial\left(x,y\right)} &=& \Lambda(\alpha)
 \label{zkompel}\\
\nonumber\\
 \left(\frac{ \partial B_{z}}{\partial\alpha}
  + \frac{\partial w_{z}}{\partial\zeta} -\frac{w_{z}}{2\rho}\,
  \frac{d\rho}{d\zeta}  \right)\,\Lambda(\alpha) &=& \lambda(\alpha) B_{z}\label{polkompel}
        \end{eqnarray}
The above relations can be shown to be valid in the following way: From
\begin{equation}
\left[ \vec\nabla\left( \frac{1}{\sqrt{\rho}}\,\frac{\partial\left(\zeta,\alpha \right)}{\partial\left(x,y\right)}
 \right) - \lambda\,\vec\nabla\alpha \right]\times\vec e_{z} = \vec 0
\end{equation}
one derives
\begin{equation}
\vec\nabla\left( \frac{1}{\sqrt{\rho}}\,\frac{\partial\left(\zeta,\alpha \right)}{\partial
 \left(x,y\right)}\right) = \lambda\,\vec\nabla\alpha\label{lcons2}
  \end{equation}
and taking the curl of both sides of the above Eq.\,(\ref{lcons2}) one gets
\begin{eqnarray}
  \vec 0\stackrel{!}{=}\vec\nabla\times\left(\lambda\,\vec\nabla\alpha \right)\quad
   \Rightarrow\quad\vec\nabla\lambda\times\vec\nabla\alpha=\vec 0\quad\Rightarrow\lambda=\lambda(\alpha)
    \end{eqnarray}
and defining
\begin{equation}
\lambda\vec\nabla\alpha:=:\vec\nabla\Lambda(\alpha)=\Lambda'(\alpha)\vec\nabla\alpha
   \end{equation}
we thus get Eq.\,(\ref{zkompel}) and Eq.\,(\ref{polkompel}). 

For the condition of a frozen in magnetic flux it is therefore necessary that
$\lambda\equiv 0$ which implies that $\Lambda(\alpha)\equiv$ constant. 
Equation (\ref{polkompel}) can be expressed in the form of an equation similar
to the telegrapher's equation for the unknown shear function $S$
\begin{eqnarray}
  -\mu_{0}\,\frac{ \partial^2 S}{\partial\alpha^2}
  + \frac{\partial^2 S}{\partial\zeta^2} - D\,\frac{\partial S}{\partial\zeta}
  -  L\frac{\partial S}{\partial\alpha} = 0\label{telegr}
      \end{eqnarray}
where
\begin{equation}
   D = \frac{1}{2}\, \frac{d\ln\rho}{d\zeta}\qquad \textrm{and} \qquad L =  
   -\mu_{0}\,\frac{d\ln\Lambda(\alpha)}{d\alpha}
\end{equation}

It is possible to rewrite equation (\ref{zkompel}) with the help of the chain 
rule
\begin{equation}
\!\!\!\!\!\!\!\!\!\!\!\frac{\partial}{\partial\zeta}\left( \frac{1}{\sqrt{\rho}}
  \frac{\partial\left(\zeta,\alpha \right)}{\partial\left(x,y\right)}\right)
    =  \left(\frac{\partial\left(\zeta,\alpha \right)}{\partial\left(x,y\right)}\right)^{-1}
     \left(\frac{\partial\alpha}{\partial y} \frac{\partial}{\partial x} - \frac{\partial\alpha}{\partial x}
     \frac{\partial}{\partial y}\right)
    \left( \frac{1}{\sqrt{\rho}}
  \frac{\partial\left(\zeta,\alpha \right)}{\partial\left(x,y\right)}\right)
  = 0 \label{linevio}
\end{equation}
which is equivalent to
\begin{equation}
\!\!\!\!\!\!\!\!\!\!\!\!\left(\frac{1}{\sqrt{\rho}}
 \frac{\partial\left(\zeta,\alpha \right)}{\partial\left(x,y\right)}
 \right)^{-1}\!\!\frac{\left(\vec B_{p}\cdot\vec\nabla\right)}{|\vec
  B_{p}|}\!\!\left(\frac{1}{\sqrt{\rho}}\frac{\partial\left(\zeta,\alpha
  \right)}{\partial\left(x,y\right)}\right)\!\equiv\!\frac{\left(\vec B_{p}
  \cdot\vec\nabla\right)}{|\vec B_{p}|}
 \ln
 \left( \frac{1}{\sqrt{\rho}}\frac{\partial\left(\zeta,\alpha\right)}{\partial\left(x,y\right)}\right)
   = 0
\end{equation}
In the following we will use Eq.\,(\ref{linevio}) to define line violation and 
therefore reconnection rates.

It should also be mentioned that 
the $z$--component of Ohm's law (Eq.\,(\ref{ohmresimhd3})) indicates,
that the asymptotical flow component within the plane around the
heliosphere, which may be inclined with respect to the asymptotical magnetic
field components within the plane, determines the constant electric field in
the invariant direction. Consequently, if we restrict to $B_{z}=0$ and 
$w_{z}=0$ and if the plasma is ideal everywhere and
the flow is field--aligned at infinity, then the flow is field--aligned
everywhere. This is due to the fact that the electric field in the 
$z$-direction is invariant. So if the electric field in the invariant
direction vanishes at infinity, the Jacobian matrix in 
Eq.\,(\ref{ohmresimhd3}) vanishes. This then implies that the flow is 
in fact
field-aligned everywhere. A similar discussion for axissymmetric equilibria 
can be found e.g. in \cite{contop}.

\subsection{Application to a plane in the vicinity of the equatorial plane
of the sun}

We set $S=0$ which results in $B_{z}=0$ and $w_{z}=0$ and take the Parker
outflow (\cite{park61}) for our investigation 
\begin{equation}
\zeta =w_{\infty }y+D_{0}\arctan\left( \frac{y}{x}\right) \,.
\label{parkerpot}
\end{equation}
The Parker flow field is defined by a potential field. In this article we assume the
auxilliary flow field and the magnetic field to be potential fields, i.e. $%
\Delta \zeta =0$ and $\Delta \alpha =0$. The stream lines of the Parker
outflow given by the potential $\zeta $ in Eq.(\ref{parkerpot}) can be seen
in Fig.\thinspace \ref{parkerflow}. In general, potential fields can be
represented by Laurent series. We define $A\equiv \phi _{m}+i\alpha $ as the
complex flux function with $\phi _{m}$, the scalar magnetic potential, as
the real, and $\alpha $ as the imaginary part of $A$. Using the assumption of
a finite number of neutral points $u_{k}$ and asymptotical boundary conditions,
$\lim\limits_{x,y\rightarrow\infty}B = B_{\infty}$, we can write: 
\[
A=B_{\infty }u+C_{0}\ln {u}+\sum_{\nu =1}^{N}\,C_{\nu }u^{-\nu
}\,\Rightarrow \,B(u_{k})=\frac{dA}{du}\,\biggr|_{u=u_{k}}=0\, .
\]
The roots of the polynomial $B(u_{k})=0$ are correlated to multipole
moments: 
\begin{eqnarray}
&&u^{N+1}+\frac{C_{0}}{B_{\infty }}u^{N}-\sum_{\nu =1}^{N}\,\frac{\nu C_{\nu
}}{B_{\infty }}u^{N-\nu }=\prod_{k=1}^{N+1}(u-u_{k})=0 \\
&\Rightarrow &C_{0}=-B_{\infty }\sum_{k=1}^{N+1}u_{k}\quad C_{\nu
}=(-1)^{\nu }\,\frac{B_{\infty }}{\nu }\,\displaystyle\sum_{\bigcup C_{\nu
+1}^{N+1}}\,\left( \prod_{u_{k}\in C_{\nu +1}^{N+1}}u_{k}\right) \,,
\label{vieta2}
\end{eqnarray}
where $1\leq \nu \leq N$. The same procedure is valid for potential flows
given by $\Delta \zeta =0$. Therefore, we can draw the conclusion, that the
topology determines the global geometry of a heliospheric model, if the
magnetic neutral point, i.e. stagnation point distribution for such a
potential magnetic field is known. The field lines, crossing a magnetic
neutral point are called magnetic separatrices. In the case of the
heliosphere this magnetopause encloses the inner region of the heliosphere,
therefore called heliopause. But on the other hand, there is also a
`hydropause', which is marked by the separatrix of the plasma flow. This separatrix
passes through the stagnation point and leads to the following questions:
what is the `heliopause'? Is it the `hydropause',\thinspace\ or is it the
`magnetopause'? If hydro-- and magnetopause are not identical, does this
include a breakdown of the magnetic connectivity with respect to the bulk plasma
flow and therefore magnetic reconnection? How large is the reconnection rate
or line violation rate in the case of potential fields with an
incompressible flow? This includes the question, whether the heliosphere is open
or closed with respect to the counterstreaming interstellar medium.

\section{Reconnection solutions}

We will analyze the magnetic reconnection processes due to the simplified
scenario we have developed in the last subsections with the help of the
critereon for line and flux conservation. This relation is given, e.g by 
\cite{vasyl} and references given therein. 
%
%
%
%
%
%
Our assumption here is, that the magnetic field looks like a Parker-spiral 
(see e.g. theory: \cite{park58} and measurements: \cite{smith}) by imitating the azimuthal component and spiral pattern of 
the field lines using a complex monopole moment representation.
Hereby, the complex monopole moment is given by $C_{0}=C_{0r}+\mathrm{{i}C_{0i}}$,
which results in the following representation of $\alpha $: 
\begin{equation}
\Im (A)=\alpha =B_{x,\infty }y-B_{y,\infty }x+C_{0r}\arctan\left( 
\frac{y}{x}\right) +C_{0i}\ln \left( \sqrt{x^{2}+y^{2}}\right) 
\end{equation}
Solving for the neutral points gives the relation between the monopole
moment on one hand and the coordinates of the neutral point and the
asymptotical components of the magnetic field on the other hand 
\begin{equation}
C_{0r}=-B_{x,\infty }x_{N}\,(-B_{y,\infty }y_{N})\quad C_{0i}=-B_{y,\infty
}x_{N}\,(-B_{x,\infty }y_{N})
\end{equation}
\begin{figure}[tbp]
\begin{center}
\includegraphics[width=6cm]{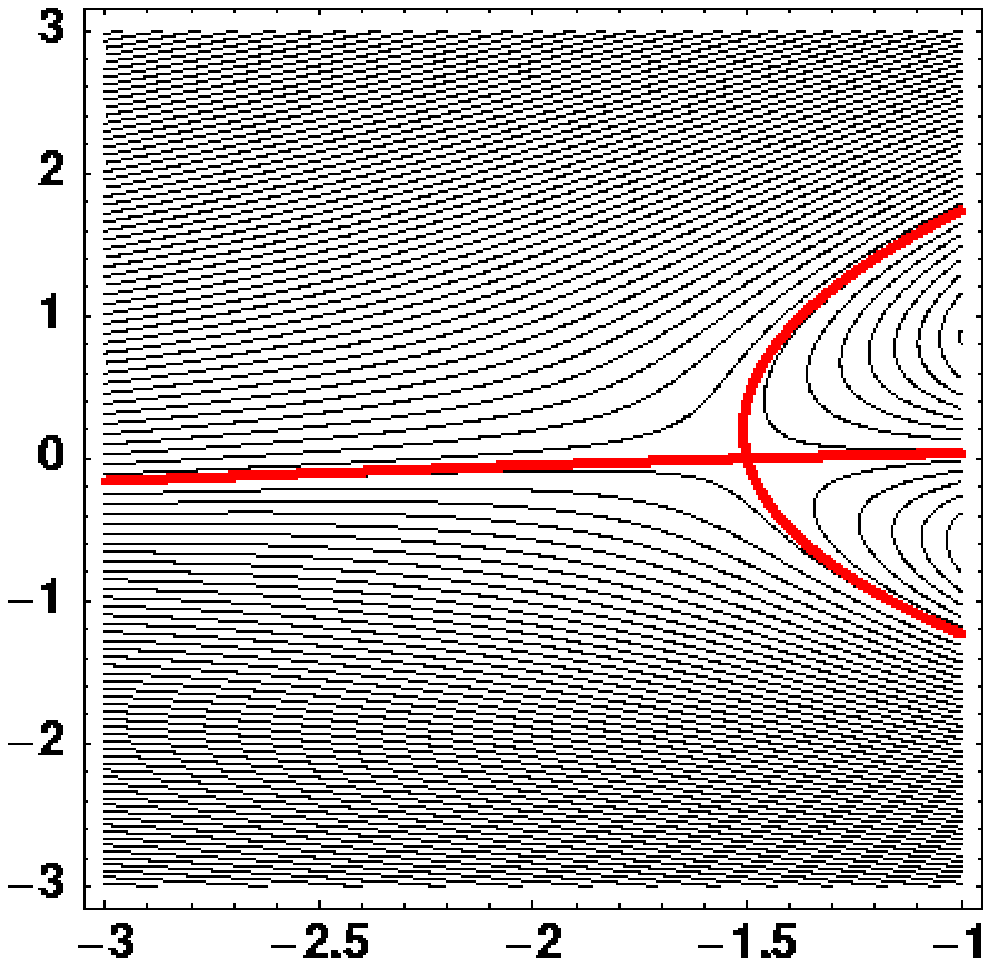}
\includegraphics[width=6cm]{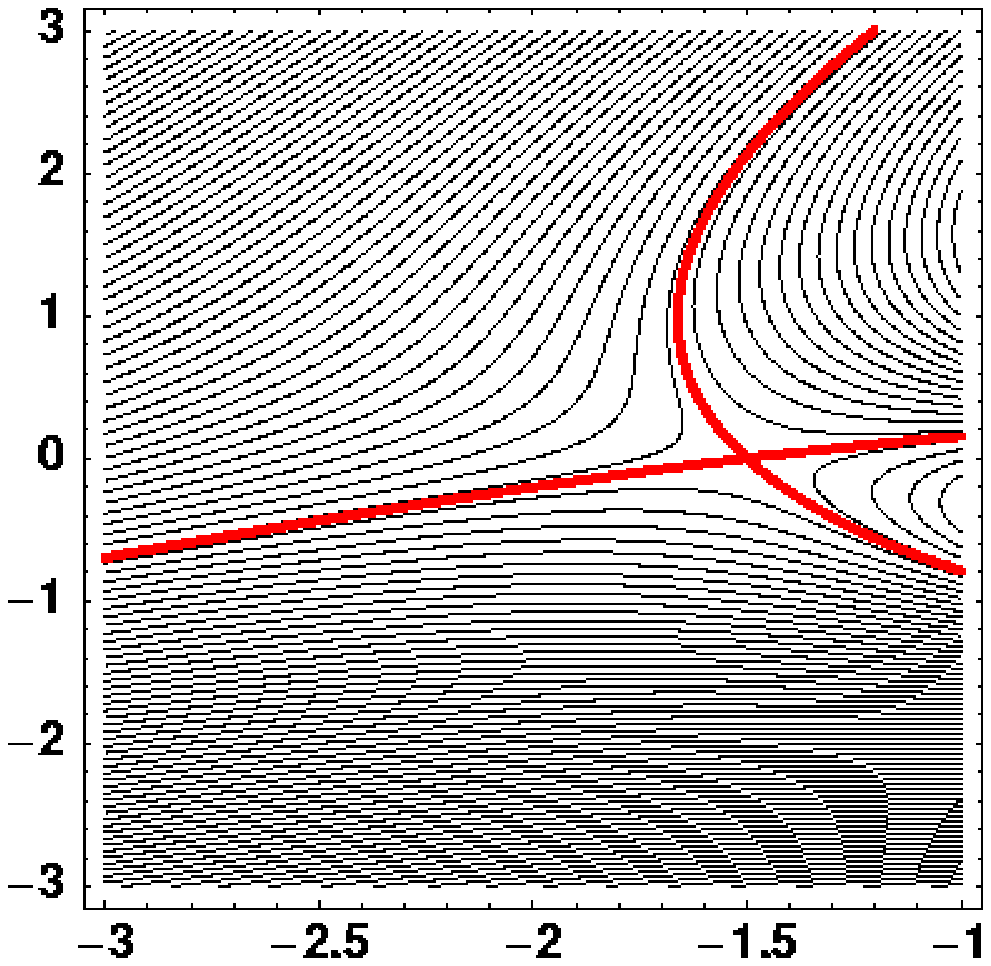}
\end{center}
\caption{The heliospheric Parker spiral like magnetic field, with $10^{\circ
}$ (left panel) and $40^{\circ }$ (right panel) tilting of the asymptotical
magnetic field with respect to the asymptotical classical Parker (out)flow;
the asymptotical Parker flow is aligned to the $x$--axis. Scale as in Fig.\,1}
\label{feldlinzedeg}
\end{figure}
How does this given potential field influence the strength and importance of
the resistivity $R_{z}$? With the help of the asymptotical boundary
conditions we get for the asymptotical electric field: 
$\lim_{x,y\rightarrow \infty }(R_{z}-E_{0}) = -E_0 = \frac{1}{\sqrt{\rho _{\infty }}}%
\,w_{\infty }B_{y\infty }$.

We calculate reconnection rates ($=$line violation rates), by using the
derivative of Ohm's law yielding 
\[
r=\frac{\partial }{\partial \zeta }\,\ln\frac{\partial \left( \zeta ,\alpha
\right) }{\partial \left( x,y\right) }
\]
where the Jacobian should be regarded as a function of $\zeta $ and $\alpha $. We use Vasyliunas' expression of line conserving flows (\cite{vasyl}),
which shows the condition for line freezing. Line freezing is fulfilled, if $
r$ vanishes everywhere. If $r$ vanishes only in the ideal region $
x,y\rightarrow \infty $, line freezing is violated and reconnection is
taking place. 
%
\begin{figure}[tbp]
\begin{center}
\includegraphics[width=6.5cm]{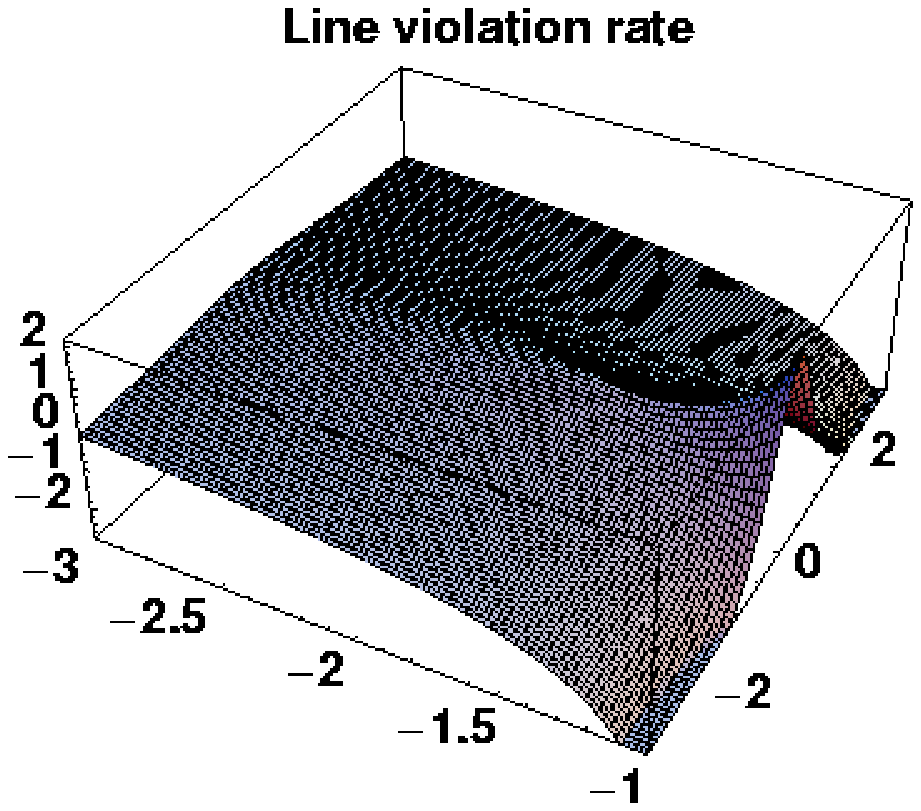} %
\includegraphics[width=6.5cm]{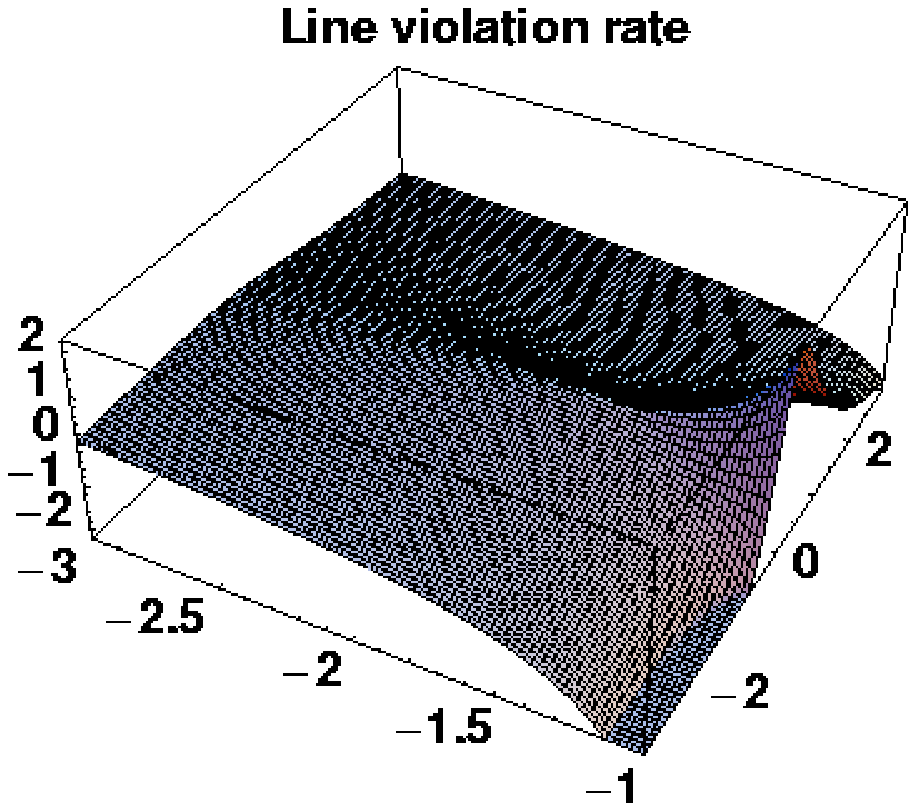}
\end{center}
\caption{The line violation rate, with $10^{\circ }$ (left panel) and $%
40^{\circ }$ (right panel) tilting of the asymptotical magnetic field with
respect to the asymptotical classical Parker (out)flow; the asymptotical
Parker flow is aligned to the $x$--axis. Scale as in Fig.\,1}
\label{ratezedeg}
\end{figure}
In this case we assume, that stagnation and magnetic neutral point are
identical to fix at least one point of these two pauses. With increasing
angle between $\vec{w}_{\infty }$ and $\vec{B}_{\infty }$, the Parker spiral
winding grows, and also the deviation of the shapes of magneto-- and
hydropause (see Figs.\thinspace \ref{parkerflow} and \ref{feldlinzedeg}).
Fig.\thinspace \ref{ratezedeg} shows that the line violation rate is much
stronger in the lower right region, where the angle $\gamma $ between the
asymptotical magnetic field and the flow is larger. Here the magnetic field
lines of the Parker--spiral have a higher curvature in the case of larger
inclination (see right panel of Fig.\thinspace \ref{ratezedeg}). In
Fig.\thinspace \ref{ratezedeg} in front of the heliopause nose there is a
bulge which shows that the reconnection rate here is larger than at other
locations in the vicinity of the stagnation point. 

\section{Conclusions}

We presented a method for calculating reconnection rates for a driven
reconnection process. Further investigations need to consider non-potential
fields to account for spontaneous processes including Ohmic heating. 
As we have shown it would be highly valuable to obtain observational 
knowledge of the location of magnetic neutral points and stagnation points 
in the heliosphere, because on the basis of that knowledge
a unique solution could be given according to our prescription.


\end{document}